\documentclass[aps,showpacs,prl,reprint,longbibliography,superscriptaddress]{revtex4-1}

\usepackage[colorlinks=true,linkcolor=blue,citecolor=blue,urlcolor=blue]{hyperref}

\usepackage{setspace} 
\usepackage{graphicx}
\usepackage{amsmath}
\usepackage{color}
\usepackage{amsmath}
\usepackage{amssymb}
\usepackage{verbatim}
\usepackage{latexsym}
\usepackage{enumerate} 
\usepackage{bm} 
\usepackage[caption=false]{subfig}
\usepackage{multirow}
\usepackage{booktabs}

\begin{document}

\title{Nuclear magnetic resonance spectroscopy as a dynamical structural probe of high pressure hydrogen}

\author{Bartomeu Monserrat} \email{bm418@cam.ac.uk}
\affiliation{TCM Group, Cavendish Laboratory, University of Cambridge, J.\,J.\,Thomson Avenue, Cambridge CB3 0HE, United Kingdom}
\author{Sharon E. Ashbrook} 
\affiliation{School of Chemistry, EaStCHEM and Centre of Magnetic Resonance, University of St Andrews, St Andrews KY16 9ST, United Kingdom}
\author{Chris J.\ Pickard} \affiliation{Department of Materials Science and Metallurgy, University of Cambridge, 27 Charles Babbage Road, Cambridge CB3 0FS, United Kingdom}
\affiliation{Advanced Institute for Materials Research, Tohoku University 2-1-1 Katahira, Aoba, Sendai, 980-8577, Japan}

\date{\today}

\begin{abstract} 
An unambiguous crystallographic structure solution for the observed phases II-VI of high pressure hydrogen does not exist due to the failure of standard structural probes at extreme pressure.
In this work we propose that nuclear magnetic resonance spectroscopy provides a complementary structural probe for high pressure hydrogen. We show that the best structural models available for phases II, III, and IV of high pressure hydrogen exhibit markedly distinct nuclear magnetic resonance spectra which could therefore be used to discriminate amongst them. As an example, we demonstrate how nuclear magnetic resonance spectroscopy could be used to establish whether phase III exhibits polymorphism. Our calculations also reveal a strong renormalisation of the nuclear magnetic resonance response in hydrogen arising from quantum fluctuations, as well as a strong isotope effect. As the experimental techniques develop, nuclear magnetic resonance spectroscopy can be expected to become a useful complementary structural probe in high pressure experiments. 
\end{abstract}

\maketitle

Hydrogen has a central place in high pressure research\,\cite{ceperley_rmp_h_he} due to its abundance in astrophysical bodies\,\cite{jupiter_model_nettelmann} and its potential to exhibit exotic properties such as phonon-mediated high temperature superconductivity\,\cite{hydrogen_superconductivity}. However, even the basic question of what are the crystal structures of the observed high pressure hydrogen phases remains elusive, largely due to the challenges faced by standard structure determination techniques under the extreme pressures reached in diamond anvil cells.

The lack of core electrons in hydrogen implies that X-rays scatter from electrons located in molecular orbitals and cannot provide information about the individual positions of protons. Nonetheless, it has been possible to perform X-ray experiments up to pressures of $190$\,GPa to establish that the hydrogen molecules are arranged in an hexagonal closed packed lattice up to those pressures\,\cite{h_xray_phase_iii,h_xray_phase_iii_2017}. 

Protons and deuterons have significant neutron scattering cross-sections, which might suggest that neutron diffraction would be an ideal probe for structural studies of high pressure hydrogen. However, the weakness of available neutron sources combined with the small samples in diamond anvil cells place practical limits to the applicability of this technique\,\cite{high_pressure_neutron_review}, and the highest pressure experiments reported in the hydrogen-deuterium system only reach $38$\,GPa\,\cite{loubeyre_d2_neutron}. 

The most widely used structural probes of high pressure hydrogen are infrared (IR) and Raman spectroscopies. Although these techniques have been extremely successful at identifying structural phase transitions in high pressure hydrogen\,\cite{h_phase_IV_eremets,gregoryanz_nature_phase_V,silvera_new_hd_phases,silvera_h2pre_science}, they only partially probe the lattice dynamics of phonons with vanishing wave vectors, and as a consequence are insufficient to determine the crystal structures of the underlying phases. 

The difficulties faced by standard structure determination techniques in the case of high pressure hydrogen suggest that alternative methods could provide valuable complementary information. Nuclear magnetic resonance (NMR) spectroscopy probes the electronic response to applied magnetic fields, for example via the chemical shielding tensor $\bm{\sigma}$ in insulators, which relates the induced magnetic field $\mathbf{B}_{\mathrm{in}}=-\bm{\sigma}\mathbf{B}_0$ at an atomic site to an applied external field $\mathbf{B}_0$. The chemical shielding tensor strongly depends on the local electronic configuration, and as a consequence encodes information about the local structure. This has led to the growing field of NMR crystallography, in which modelling and experiment are combined to solve crystal structures\,\cite{nmr_crystallography_example}. 

There has been long-term interest in combining NMR spectroscopy with diamond anvil cell high pressure experiments, and reports include studies of ortho-para hydrogen conversion up to $12.8$\,GPa\,\cite{silvera_nmr_ortho_para} and proton diffusion up to $6.8$\,GPa\,\cite{h2_nmr_diffusion}. Experiments are challenging due to a variety of factors, including weak signals, low resonator sensitivities, and poor access to the samples in the anvil cells, which have traditionally limited high pressure NMR spectroscopy experiments to a maximum of a few tens of GPa\,\cite{review_nmr_gpa}. This situation has recently changed with the ground-breaking developments of Meier and co-workers, who exploiting magnetic flux tailoring Lenz lenses to amplify the magnetic field at the sample have measured hydrogen NMR spectra of paraffin up to $72$\,GPa\,\cite{lenz_lenses_nmr_high_pressure} and of FeH up to $200$\,GPa\,\cite{meier_nmr_feh_200GPa}. The state-of-the-art of high pressure NMR spectroscopy has recently been reviewed in Refs.\,\cite{Meier_high_pressure_nmr_review1,Meier_high_pressure_nmr_review2}. These recent developments pave the path towards NMR spectroscopy experiments in the pressure range relevant for high pressure hydrogen phases II-VI. Phase II of hydrogen first appears at pressures in the range $73$--$110$\,GPa\,\cite{silvera_phase_ii,hemley_phase_ii_iii,phase_ii_gregoryanz}, and that of deuterium at even lower pressures of about $25$\,GPa\,\cite{silvera_phase_ii_deuterium}. Phase III appears in the pressure range $150$--$170$\,GPa\,\cite{h_phase_III}, and phase IV in the pressure range $200$--$220$\,GPa\,\cite{h_phase_IV_gregoryanz}. These are all within reach of the latest high pressure NMR spectroscopy experiments. 

In this work we study the potential of NMR spectroscopy to probe the structure of high pressure hydrogen phases. In particular, we show that NMR chemical shieldings are markedly distinct between the various theoretical structural models proposed for hydrogen insulating phases II, III, and IV, and could therefore be used to assign the appropriate structure to these phases. We also expect that NMR spectroscopy could play a major role in other high pressure systems, including the hydrides which have recently been shown to exhibit some of the exotic properties initially proposed for pure hydrogen\,\cite{ashcroft_superconducting_hydrides,Duan_reports_h3s,Drozdov_2015_superconductor}, and also compounds containing heavier elements that are present in the interiors of gas and rocky planets.

We consider six hydrogen structures. The $P2_1c$\,\cite{nature_physics_h} and $P6_3m$\,\cite{nature_physics_h} structures are candidates for phase II, which exists at low temperatures and at pressures up to around $150$\,GPa. The $C2/c$\,\cite{nature_physics_h} and $P6_122$\,\cite{hexagonal_phase_III_monserrat} structures are candidates for phase III, which exists in a wide pressure range above about $150$\,GPa. The $Pc$\,\cite{phase_iv_prb} and $Pca2_1$\,\cite{Pca21_phase_V} structures are candidates for phases IV/V, which appear around room temperature and at pressures higher than $220$\,GPa. All these structures have low enthalpies and Gibbs free energies\,\cite{prl_dissociation_hydrogen,hydrogen_nature_communications,h_dissociation_morales}, rendering them realistic candidates for the experimentally observed phases. Furthermore, several of their structural\,\cite{hexagonal_phase_III_monserrat} and spectroscopic\,\cite{md_raman_ackland,ackland_finite_temp_raman_h,md_ramand_ir_azadi,ackland_finite_temp_ir_h} features are consistent with corresponding experimental observations, although there are some outstanding discrepancies, particularly regarding the nature of their band gaps\,\cite{ceperley_h_elph_coupling,azadi_h_metal_dmc_static,azadi_gap_pimd}. Available experimental and theoretical data is insufficient to unambiguously identify the correct structure associated with the observed experimental phases: for example both $C2/c$ and $P6_122$ structures exhibit IR and Raman spectra consistent with phase III, and the $Pc$ and $Pca2_1$ structures consistent with phase IV, while their relative enthalpies are very similar.

To explore the possibility that NMR spectroscopy might provide complementary structural information, we have performed first principles density functional theory (DFT) calculations using the {\sc castep} package\,\cite{CASTEP}. We have optimised the volumes and internal coordinates of the six structures by minimising their enthalpies at pressures in the range $150$--$250$\,GPa, and using five different approaches, namely the local density approximation (LDA)\,\cite{PhysRevLett.45.566,PhysRevB.45.13244}, the Perdew-Burke-Ernzerhof (PBE) approximation\,\cite{PhysRevLett.77.3865}, and the Becke-Lee-Yang-Parr (BLYP) approximation\,\cite{blyp_exchange} the PBE approximation with the Tkatchenko-Scheffler van der Waals correction (PBE+TS)\,\cite{ts_vdW}, and the hybrid Heyd-Scuseria-Ernzerhof (HSE) approximation,\cite{hse03_functional,hse06_functional} to describe the exchange-correlation functional. We have then calculated the chemical shielding tensor using the gauge-including projector augmented waves (GIPAW) theory\,\cite{gipaw,gipaw_ultrasoft,gipaw_review} as implemented in {\sc castep}. 

\begin{figure}
\centering
\includegraphics[scale=0.37]{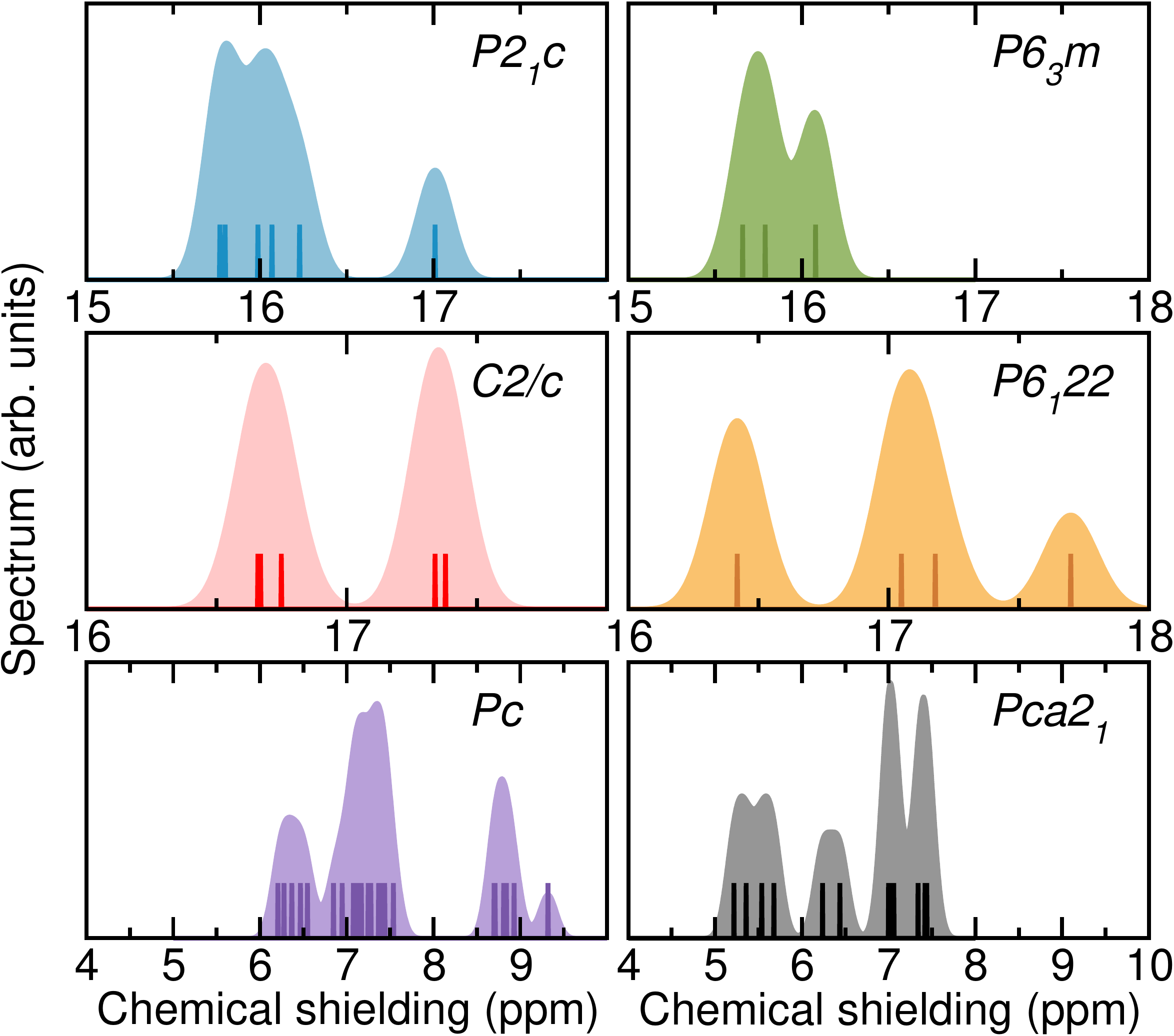}
\caption{Isotropic chemical shieldings calculated using the PBE functional at $150$\,GPa for the $P2_1c$, $P6_3m$, $C2/c$, and $P6_122$ structures, and at $250$\,GPa for the $Pc$ and $Pca2_1$ structures. The vertical lines show the precise location of the calculated shieldings, while the wider spectra are the result of a convolution with a Gaussian function of $0.1$\,ppm width.} 
\label{fig:static-spectra-pbe}
\end{figure}

The main result of our work is presented in Fig.\,\ref{fig:static-spectra-pbe}, showing the isotropic chemical shieldings $\frac{1}{3}\mathrm{Tr}(\bm{\sigma})$ of the six considered structures calculated using the PBE functional. We also show the corresponding maximum and minimum absolute shielding anisotropies $\sigma_{\mathrm{SA}}=|\sigma_{33}-\frac{1}{2}(\sigma_{11}+\sigma_{22})|$ in Table\,\ref{tab:sa}. Isotropic lineshapes could be obtained experimentally using magic angle spinning (MAS)\,\cite{magic_angle_spinning}, but we note that the precise parameters measured depend on the type of experiment performed, and we also provide magres files containing the full chemical shielding tensors with the Supplemental Material to serve as the basis for further processing for direct comparison with future experiments.

\begin{table}
  \setlength{\tabcolsep}{6pt} 
  \caption{Shielding anisotropies calculated using the PBE functional at $150$\,GPa for the $P2_1c$, $P6_3m$, $C2/c$, and $P6_122$ structures, and at $250$\,GPa for the $Pc$ and $Pca2_1$ structures.}
  \label{tab:sa}
  \begin{ruledtabular}
  \begin{tabular}{ccc}
   Structure & \multicolumn{2}{c}{Shielding anisotropy (ppm)}  \\
             & maximum value & minimum value \\
  \hline
  $P2_1c$  & $7.78$  & $3.87$  \\
  $P6_3m$  & $9.22$ & $4.05$  \\
  $C2/c$   & $4.50$  & $1.86$  \\
  $P6_122$ & $4.49$  & $1.50$  \\
  $Pc$     & $6.56$  & $2.93$ \\
  $Pca2_1$ & $6.18$  & $0.56$ \\
\end{tabular}
\end{ruledtabular}
\end{table}

Comparing the NMR spectra of the phase II candidate structures at $150$\,GPa in Fig.\,\ref{fig:static-spectra-pbe}, we observe that the $P2_1c$ structure exhibits two closely spaced peaks centred around $16$\,ppm, and a separate narrower peak at around $17$\,ppm. There are six distinct atomic sites (each with four atoms) in the $P2_1c$ structure, but five of these give similar chemical shieldings and result in the broad double peak, while the sixth site gives the lone peak at $17$\,ppm. By contrast, the $P6_3m$ structure only exhibits two closely spaced peaks centred at $16$\,ppm, which are formed by three distinct chemical shieldings arising from the three independent atomic sites (two of these with six atoms each, and one with four atoms) in this structure. Analogous calculations using the LDA and BLYP functionals (see Supplemental Material) consistently show a significantly broader spectrum for the $P2_1c$ structure compared with the $P6_3m$ structure. The different range and number of peaks between the two structures indicate that NMR spectroscopy could provide a promising tool to solve the structure of hydrogen phase II. It could also shed light on the potential polymorphism in phase II of deuterium\,\cite{goncharov_phase_ii_polymorphism,phase_ii_gregoryanz}. 

Comparing the NMR spectra of the phase III candidate structures at $150$\,GPa in Fig.\,\ref{fig:static-spectra-pbe}, we observe that the $C2/c$ structure exhibits two peaks centred at $16.70$\,ppm and $17.35$\,ppm, which is a markedly different spectrum to that of the $P6_122$ structure with three peaks at $16.40$\,ppm, $17.10$\,ppm, and $17.70$\,ppm. There are six distinct atomic sites (each with four atoms) in the $C2/c$ structure, which lead to six distinct chemical shieldings, but these cluster into two groups of three yielding the two observed peaks at the $0.1$\,ppm resolution. There are also six distinct sites (each with six atoms) in the $P6_122$ structure, yielding four distinct chemical shieldings (two pairs of sites yield the same shielding), two of which are similar and lead to the central peak, and the other two lead to peaks at the highest and lowest shieldings. We note that the peak at the lowest shielding has greater intensity than that at the highest shielding because the former arises from $12$ atomic sites, while the latter from six. This clear difference between the NMR spectra of the $C2/c$ and $P6_122$ structures could prove critical in understanding the properties of phase III. Theoretically, the $P6_122$ structure has lower Gibbs free energy at temperatures in the range $0$--$300$\,K and at pressures below $200$\,GPa and the $C2/c$ structure above that pressure, suggesting the possibility of polymorphism in phase III\,\cite{hexagonal_phase_III_monserrat}. However, available experimental data is insufficient to confirm or reject this hypothesis, because both structures exhibit essentially indistinguishable IR and Raman spectra, consistent with experiment, and X-ray diffraction data is only available below $200$\,GPa\,\cite{h_xray_phase_iii,h_xray_phase_iii_2017}, suggesting a hexagonal structure consistent with $P6_122$. The distinct NMR spectra exhibited by these two structures could unambiguously resolve the question of polymorphism in phase III. We note that the $2$-peak spectrum of the $C2/c$ structure and the $3$-peak spectrum of the $P6_122$ structure is maintained if the calculations are performed using the LDA, PBE+TS, or HSE functionals, while the peaks move closer together when using the BLYP functional, but the latter describes the hydrogen bond more poorly (see Supplemental Material).

Comparing the NMR spectra of the phase IV/V candidate structures at $250$\,GPa in Fig.\,\ref{fig:static-spectra-pbe}, we observe that the $Pc$ structure exhibits a broad double peak between $6.00$ and $7.75$\,ppm and a narrower double peak between $8.50$ and $9.50$\,ppm. There are $24$ distinct atomic sites (each with two atoms) in the $Pc$ structure, and each leads to a different value for the chemical shielding which results in the broad peaks observed in Fig.\,\ref{fig:static-spectra-pbe}. The NMR spectra of the $Pca2_1$ structure exhibits a continuum of peaks spanning the range from $5.00$ to $7.75$\,ppm. These peaks arise from $12$ distinct atomic sites (each with four atoms), a smaller number compared to the $Pc$ structure. The only available experimental information on phases IV/V is based on their Raman and IR spectra\,\cite{h_phase_IV_eremets,gregoryanz_nature_phase_V}, which are largely consistent with those of both $Pc$ and $Pca2_1$ structures\,\cite{md_raman_ackland,Pca21_phase_V}. Again, our results show that NMR spectroscopy could provide further evidence to disentangle the underlying structure of these phases of hydrogen. We note that using the LDA functional leads to a similar picture to that shown in Fig.\,\ref{fig:static-spectra-pbe}, in which the $Pc$ NMR spectrum spans a wider range of chemical shieldings than the spectrum of the $Pca2_1$ structure. The comparison using the BLYP functional is not possible because the $Pca2_1$ structure exhibits metallic behaviour in that case (see Supplemental Material).

\begin{figure}
\centering
\includegraphics[scale=0.37]{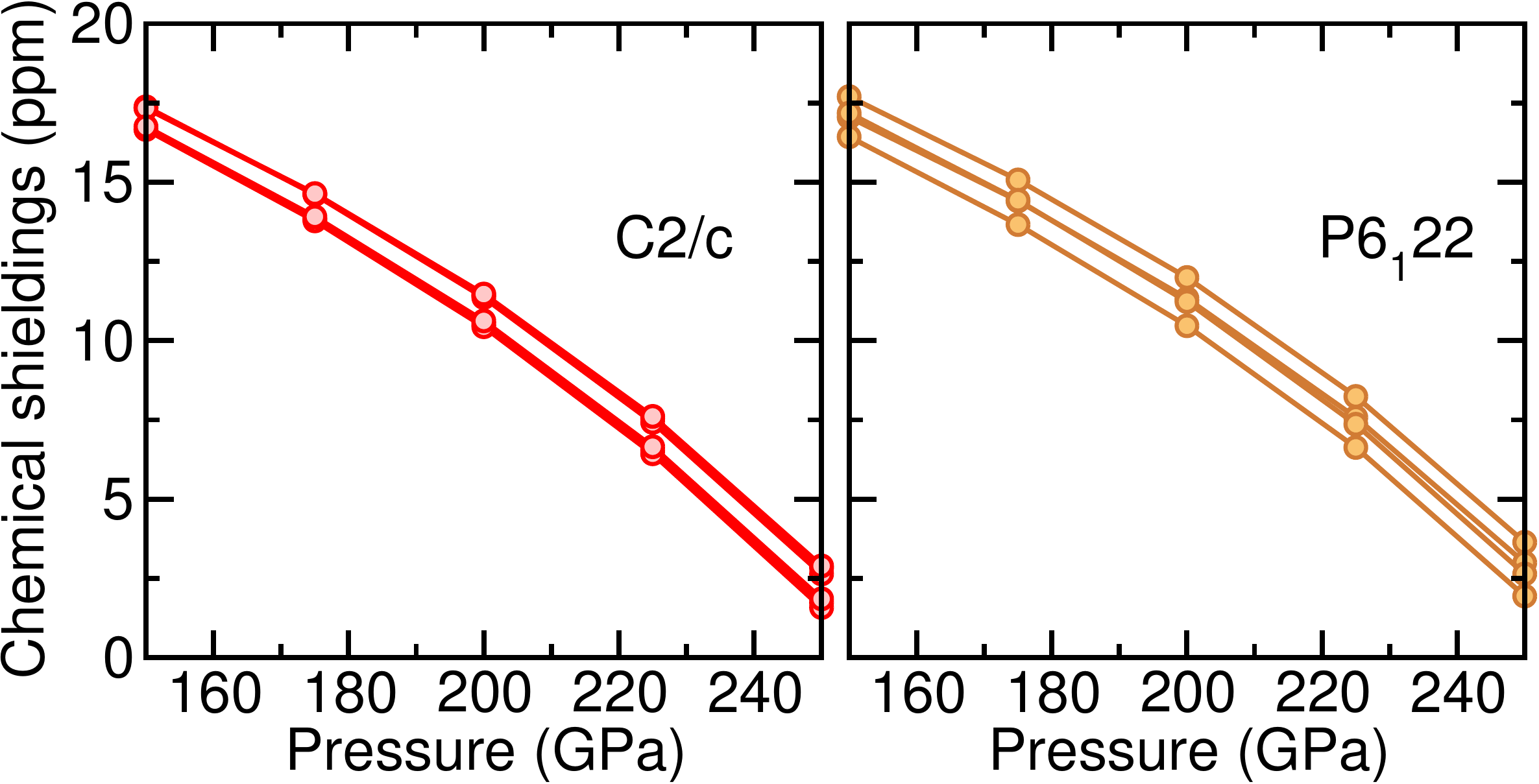}
\caption{Isotropic chemical shieldings of the $C2/c$ and $P6_122$ structures calculated using the PBE functional as a function of pressure.} 
\label{fig:static-pressure-pbe}
\end{figure}

We have repeated the chemical shielding tensor calculations of the phase III candidate $C2/c$ and $P6_122$ structures as a function of pressure in the range $150$ to $250$\,GPa, and the results are depicted in Fig.\,\ref{fig:static-pressure-pbe}. There is a relatively strong shift of the NMR peak positions as a function of pressure from around $17$\,ppm at $150$\,GPa to about $3$\,ppm at $250$\,GPa. Nonetheless, the relative shieldings of the spectra depicted in Fig.\,\ref{fig:static-spectra-pbe} for $150$\,GPa are maintained at all the considered pressures, suggesting that experiments at any pressure in which phase III is observed should be sufficient to fully explore the relation between the NMR spectra and the underlying structures. We also remark that NMR spectroscopy experiments do not measure absolute shieldings, but instead relative shifts with respect to some reference structure, and therefore the relevant quantity in our predictions is the relative position of the peaks, not their absolute value.


All calculations reported so far have been performed at the static lattice level of theory. However, hydrogen is the lightest atom and quantum zero point motion is known to strongly renormalise its energetic\,\cite{hydrogen_nature_communications}, structural\,\cite{morales_pimd_h_structure}, optical\,\cite{ceperley_h_elph_coupling,azadi_gap_pimd}, and vibrational\,\cite{md_raman_ackland,md_ramand_ir_azadi} properties. We have therefore performed chemical shielding tensor calculations for the $C2/c$ and $P6_122$ structures including the effects of quantum zero-point motion. The initial step is the calculation of the lattice dynamics at the harmonic level of theory, which we have performed using DFT as implemented in the {\sc castep} package, and using the finite displacements method in conjunction with nondiagonal supercells\,\cite{non_diagonal}. Anharmonic lattice dynamics contributions are known to be very important for calculating the relative energy of different hydrogen structures, but this is mostly due to the small energy differences between structures rather than a significant anharmonic energy, which is only about $5$\% of the harmonic energy\,\cite{hydrogen_nature_communications}. We have therefore neglected anharmonic terms in our lattice dynamics calculations. 

\begin{figure}
\centering
\includegraphics[scale=0.37]{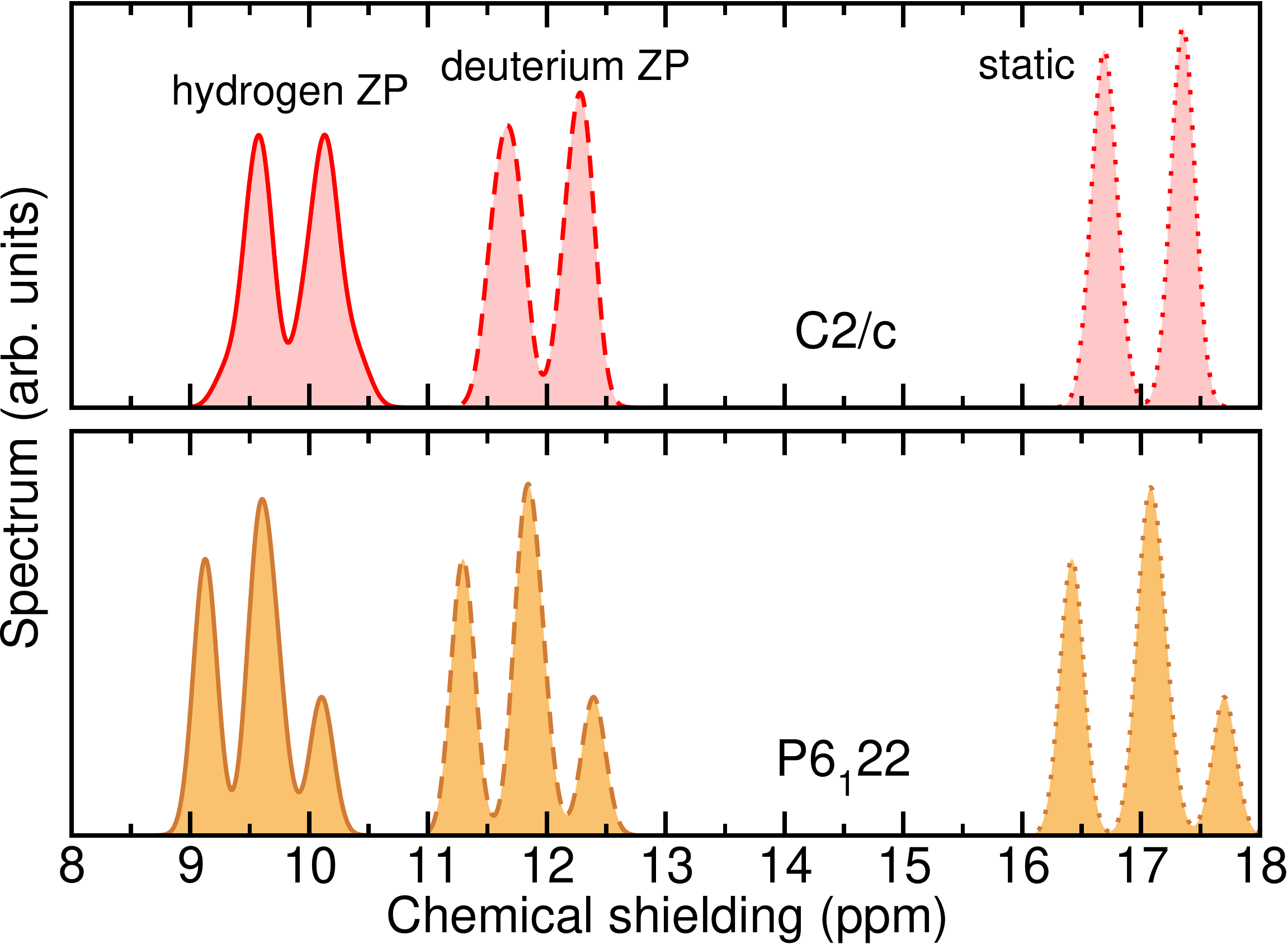}
\caption{Isotropic chemical shieldings of $C2/c$ and $P6_122$ calculated using the PBE functional at $150$\,GPa. The dotted line distributions show the static lattice results, and the dashed and solid line distributions show the results including quantum zero-point (ZP) motion for deuterium and hydrogen, respectively. The results have been convoluted with a Gaussian function of $0.1$\,ppm width.}
\label{fig:zp}
\end{figure}

We then calculate the zero point quantum renormalisation of the chemical shielding tensor using the stochastic approach described in Refs.\,\cite{pickard_original_nmr_vibrational,monserrat_nmr_tdep_original,errea_stochastic_nmr}, and accelerating the sampling by exploiting thermal lines\,\cite{thermal_lines,jpcm_elph_review_monserrat}. 
The isotropic chemical shieldings of the $C2/c$ and $P6_122$ structures calculated with the inclusion of quantum zero-point motion are depicted in Fig.\,\ref{fig:zp}. Zero-point motion leads to a strong renormalisation of the peak positions of about $7$\,ppm, confirming that quantum fluctuations are strong in hydrogen. However, the number and relative position of the peaks in the spectra of the two structures remains similar to that of the static lattice results. This suggests that, although quantum zero point motion strongly renormalises the electronic response to magnetic fields in hydrogen, the results depicted in Fig.\,\ref{fig:static-spectra-pbe} showing the differences in the NMR spectra between different high pressure hydrogen candidate structures remain valid. 

Figure \ref{fig:zp} also shows the quantum zero-point renormalisation of the chemical shielding corresponding to the deuterium isotope. As a consequence of the large mass difference between hydrogen and deuterium, the zero-temperature chemical shieldings exhibit a strong isotope effect. This suggests that NMR spectroscopy could also be used to explore the isotopic composition of high pressure hydrogen phases. In addition, we note that the deuterium nucleus is quadrupolar and it will therefore be subject to an electric field gradient (relevant parameters listed in the Supplemental Material). This implies that in experiments without MAS, the spectral lines will be broadened by the quadrupolar interaction in deuterium. 

Our results indicate that NMR spectroscopy is a promising technique for elucidating the structures of the observed phases of high pressure hydrogen. Direct observation of the calculated isotropic chemical shieldings would require MAS of a proton sample in a diamond anvil cell. Although MAS has been successfully combined with diamond anvil cells, the large rotor speeds needed for proton signals might pose technical challenges due to the small size rotors required. Alternatively, MAS experiments on deuterium samples with residual protons would allow slower spinning and thus larger rotors, simplifying experiments. We note that good sensitivities are obtained in samples with proton concentrations as low as $1$\%\,\cite{proton_nmr_in_deuterium}. Finally, the technically simpler experiments would be static, but in those the presence of anisotropies implies that measurements should exploit differences in proton chemical shielding anisotropies or deuterium quadrupolar couplings rather than isotropic chemical shieldings. Magres files containing the full chemical shielding tensors are published with the Supplemental Material.

In conclusion, we propose that nuclear magnetic resonance could become a valuable probe to identify the underlying structures of the observed high pressure phases of solid hydrogen. We have shown that isotropic NMR spectra would be particularly useful as the chemical shieldings are markedly different between the different structural models available for hydrogen phases II, III, and IV/V. For example, they could be used to resolve questions such as the potential polymorphism of phase III or the precise structural sequence in phases IV/V. Our results pave the path for the use of NMR spectroscopy in structure determination in other high pressure systems, both for hydrogen-rich compounds or for other compounds containing heavier atoms.

\acknowledgments

B.M. acknowledges support from the Winton Programme for the Physics of Sustainability, and from Robinson College, Cambridge, and the Cambridge Philosophical Society for a Henslow Research Fellowship. S.E.A. and C.J.P. are supported by the Royal Society through a Royal Society Wolfson Research Merit award. Part of the calculations were performed using the Archer facility of the UK's national high-performance computing service (for which access was obtained via the UKCP consortium [EP/P022596/1]).

\bibliography{/Users/bartomeumonserrat/Documents/research/papers/references/h-nmr,/Users/bartomeumonserrat/Documents/research/papers/references/long,/Users/bartomeumonserrat/Documents/research/papers/references/high_pressure,/Users/bartomeumonserrat/Documents/research/papers/references/nmr}

\onecolumngrid
\clearpage
\begin{center}
\textbf{\large Supplemental Material for ``Nuclear magnetic resonance spectroscopy as a dynamical structural probe of high pressure hydrogen'' }
\end{center}
\setcounter{equation}{0}
\setcounter{figure}{0}
\setcounter{table}{0}
\setcounter{page}{1}
\makeatletter
\renewcommand{\theequation}{S\arabic{equation}}
\renewcommand{\thefigure}{S\arabic{figure}}
\renewcommand{\bibnumfmt}[1]{[S#1]}
\renewcommand{\citenumfont}[1]{S#1}

\section{Computational details}

Density functional theory (DFT) calculations have been performed with the plane-wave pseudopotential {\sc castep} code\,\cite{CASTEP} using the LDA\,\cite{PhysRevLett.45.566,PhysRevB.45.13244}, PBE\,\cite{PhysRevLett.77.3865}, BLYP\,\cite{blyp_exchange}, PBE corrected with the Tkatchenko-Scheffler van der Waals scheme (PBE+TS)\,\cite{ts_vdW} and the hybrid HSE\,\cite{hse03_functional,hse06_functional} exchange-correlation functionals. The reported calculations use cut-off energies of $1000$\,eV and electronic Brillouin zone $\mathbf{k}$-point grids of density $2\pi\times0.025$~\AA$^{-1}$ for all exchange-correlation functionals, except the HSE calculations which use a cut-off energy of $800$\,eV and a $2\pi\times0.050$~\AA$^{-1}$ $\mathbf{k}$-point grid.

\section{Exchange-correlation functional comparison}

\subsection{Semilocal exchange-correlation functionals}

\begin{figure}
\subfloat[][LDA.]{
  \includegraphics[scale=0.36]{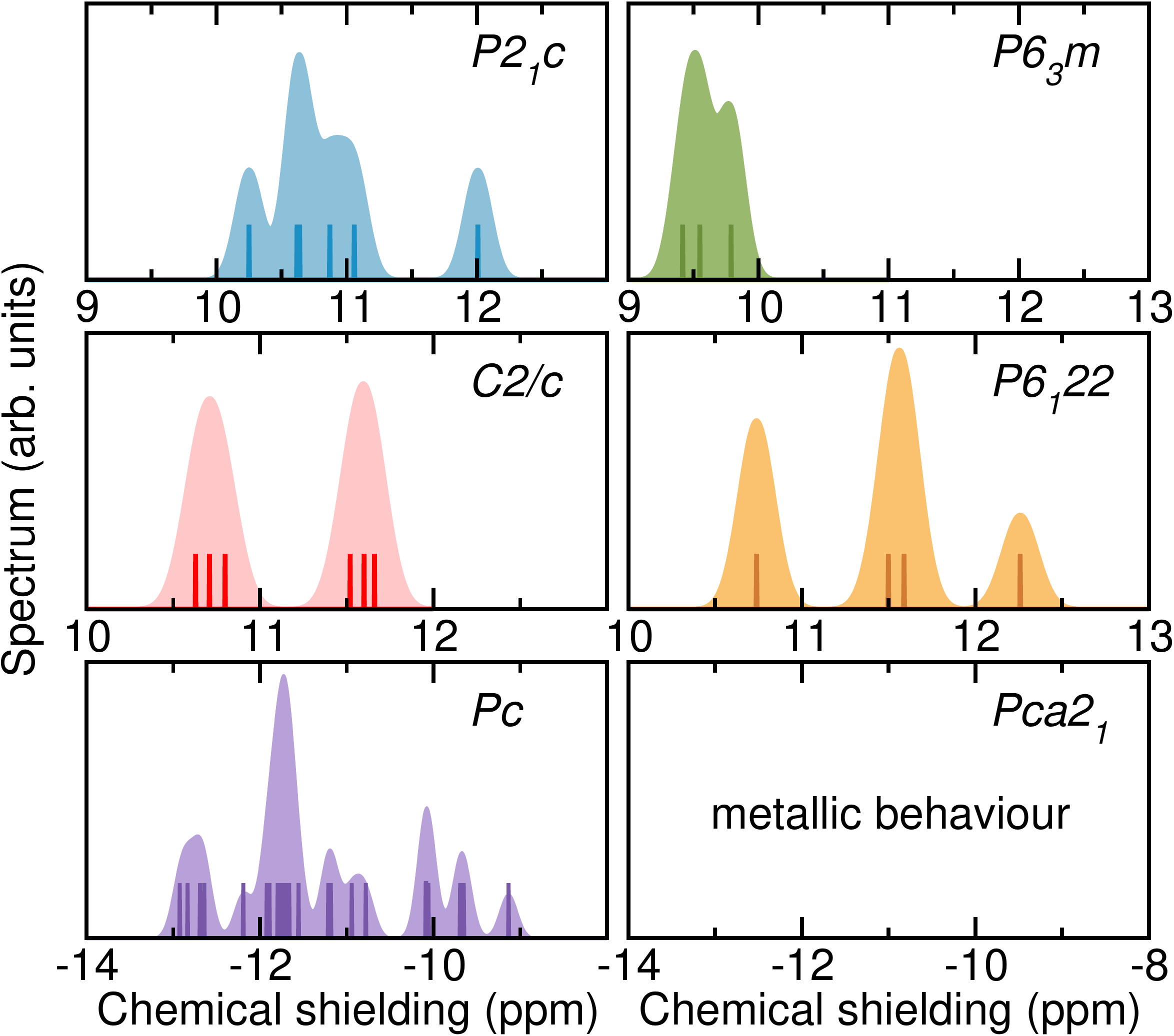}
\label{subfig:fe_static}}
\subfloat[][BLYP.]{
  \includegraphics[scale=0.36]{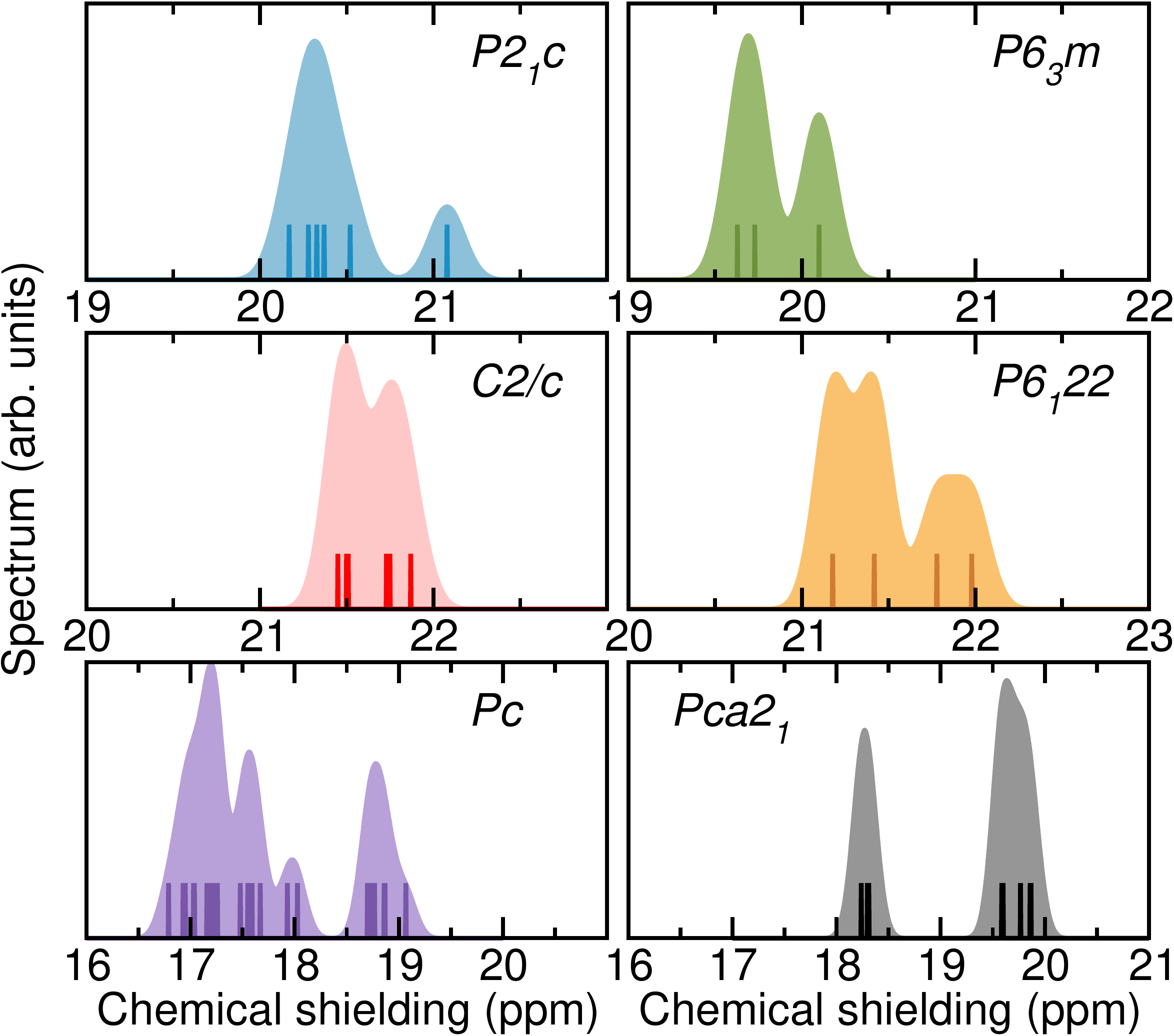}
\label{subfig:fe_gibbs}}
  \caption{Isotropic chemical shieldings calculated using the LDA and BLYP functionals at $150$\,GPa for the $P2_1c$, $P6_3m$, $C2/c$, and $P6_122$ structures, and at $250$\,GPa for $Pc$ and $Pca2_1$ structures. The vertical lines show the precise location of the calculated shifts, while the wider spectra are the result of a convolution with a Gaussian function of $0.1$\,ppm width.}
      \label{fig:static-spectra}
\end{figure}

The NMR spectra of the six structures considered in the main text are reported in Fig.\,\ref{fig:static-spectra} using the LDA and BLYP functionals. Consistently with the main text, the shieldings for $P2_1c$, $P6_3m$, $C2/c$, and $P6_122$, are calculated at $150$\,GPa, and those for $Pc$ and $Pca2_1$ at $250$\,GPa. The absolute value of the chemical shifts differs significantly depending on the functional used, but NMR spectra are not recorded in absolute terms experimentally, so the important quantity is the relative position of the peaks.

The phase II candidate structures exhibit similar relative spectra with all functionals. $P2_1c$ has a broad range of chemical shifts: $2$\,ppm for PBE (split into two main peaks at the $0.1$\,ppm resolution); $2.5$\,ppm for LDA (split into two main peaks at the $0.1$\,ppm resolution); and $2$\,ppm in BLYP (again split into two main peaks at the $0.1$\,ppm resolution). By contrast, $P6_3m$ has a single narrower peak of a width narrower than $1$\,ppm using all three functionals.

The phase III candidate structures show relatively consistent spectra between PBE and LDA, clearly exhibiting a $2$-peak structure for $C2/c$ and a $3$-peak structure for $P6_122$. Using the BLYP functional, the two peaks for $C2/c$ merge into a broader peak, and so do the three peaks of the $P6_122$ structure. However, we remark that the BLYP functional predicts H$_2$ bond lengths that are too short compared to experiment\,\cite{Pca21_phase_V}, and therefore the latter functional is expected to be least reliable.

The phase IV candidate structures are the largest with $48$ atoms per primitive cell, and as a consequence exhibit the more complex chemical shift spectra. The $Pc$ structure consistently shows a broad spectrum spanning $3$--$4$\,ppm. The $Pca2_1$ structure shows a narrower spectrum at the PBE and BLYP levels of theory, spanning about $2$--$2.5$\,ppm. The NMR spectrum of $Pca2_1$ could not be obtained using the LDA functional because the structure exhibits metallic behaviour.

\subsection{Exchange-correlation functionals with non-local correlations}

\begin{figure}
  \includegraphics[scale=0.36]{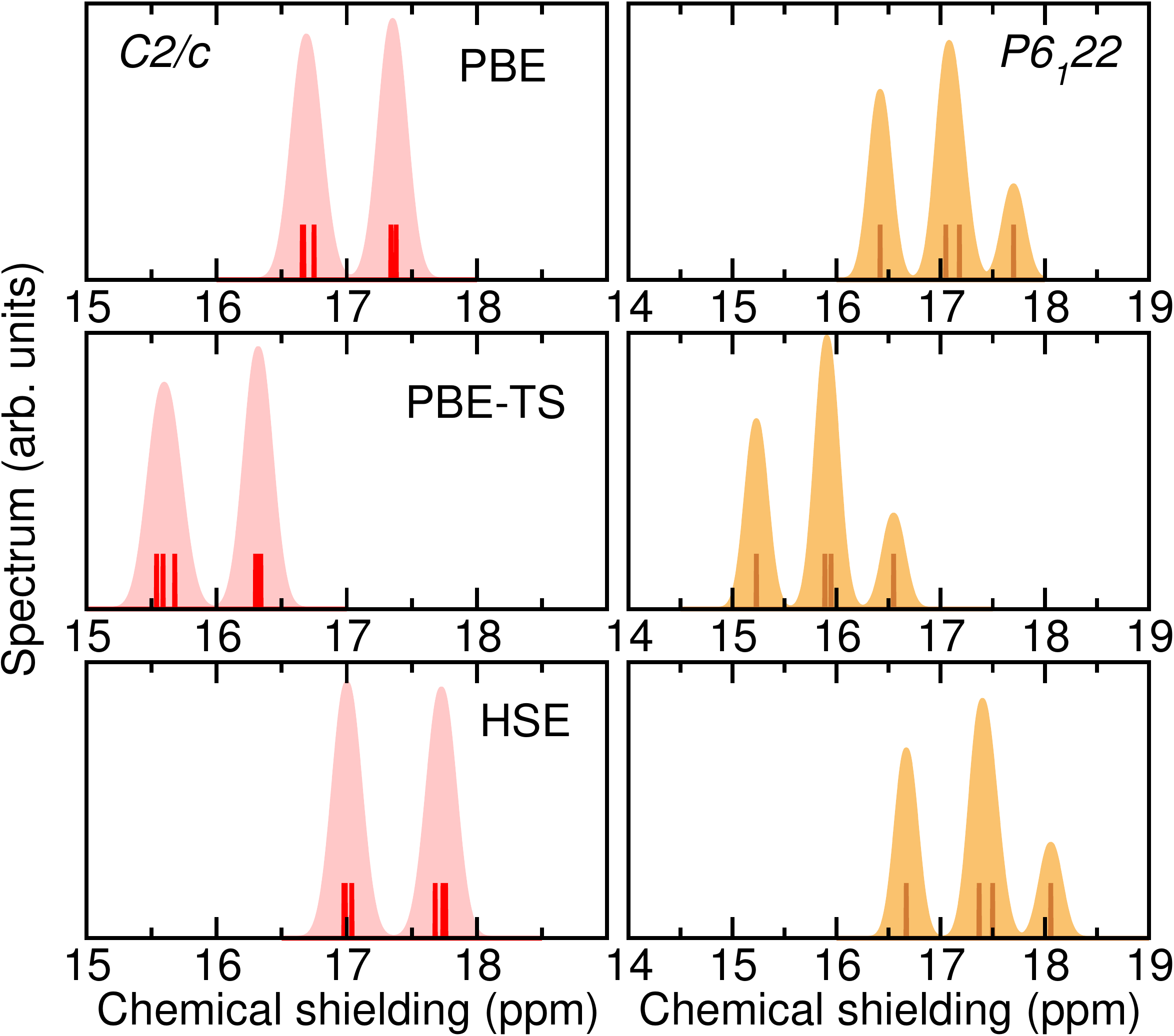}
  \caption{Isotropic chemical shieldings calculated using structures relaxed within the PBE, PBE+TS, and HSE functionals at $150$\,GPa for the $C2/c$ and $P6_122$ structures. The vertical lines show the precise location of the calculated shifts, while the wider spectra are the result of a convolution with a Gaussian function of $0.1$\,ppm width.}
      \label{fig:vdw-hse}
\end{figure}

The NMR spectra for the $C2/c$ and the $P6_122$ structures at $150$\,GPa are reported in Fig.\,\ref{fig:vdw-hse}. The results correspond to calculations using the PBE functional but with structures whose geometry optimization has been performed using the PBE functional (top row, repeated here from the main text for completeness), the PBE+TS functional (middle row), and the HSE functional (bottom row). We note that the largest effect of changing the exchange-correlation functional is in the geometry rather than in the calculation of the NMR spectrum itself, as confirmed by the close agreement between NMR spectra calculated using the LDA and PBE functionals on a PBE relaxed structure, while the difference in the LDA NMR spectra between LDA and PBE relaxed structures is larger.

The results depicted in Fig.\,\ref{fig:vdw-hse} confirm that our conclusions are also robust when the effects of non-local correlations included either via the van der Waals interaction or the non-local exchange interaction.

\section{Quantum zero-point vibrational calculations}

The chemical shielding tensor renormalised by the presence of quantum zero-point vibrations can be calculated as:
\begin{equation}
\bm{\sigma}_{\mathrm{ZP}}=\int d\mathbf{u}|\chi(\mathbf{u})|^2\bm{\sigma}(\mathbf{u}), \label{eq:vib_exp}
\end{equation}
where $\mathbf{u}$ is a vector whose elements $\{u_{\mathbf{q}\nu}\}$ are the normal mode amplitudes describing harmonic phonons and labelled by wavevector $\mathbf{q}$ and branch $\nu$, and $\chi(\mathbf{u})=\prod_{\mathbf{q}\nu}\phi_{\mathbf{q}\nu}(u_{\mathbf{q}\nu})$ is the harmonic wave function. The harmonic wave function is a product over Gaussian functions
\begin{equation}
\phi_{\mathbf{q}\nu}(u_{\mathbf{q}\nu})=\frac{1}{\sqrt{2\pi s^2_{\mathbf{q}\nu}}}\exp\left(-\frac{u^2_{\mathbf{q}\nu}}{2s^2_{\mathbf{q}\nu}}\right) \label{eq:gaussian}
\end{equation}
with $s^2_{\mathbf{q}\nu}=(2\omega_{\mathbf{q}\nu})^{-1}$ for mode frequency $\omega_{\mathbf{q}\nu}$. Equation\,(\ref{eq:vib_exp}) is evaluated stochastically as
\begin{equation}
\bm{\sigma}_{\mathrm{ZP}}\simeq\frac{1}{N}\sum_{i=1}^N\bm{\sigma}(\mathbf{u}_i), \label{eq:mc}
\end{equation}
using $N$ configurations $\mathbf{u}_i$ distributed according to $|\chi(\mathbf{u})|^2$.

\begin{figure}
  \includegraphics[scale=0.36]{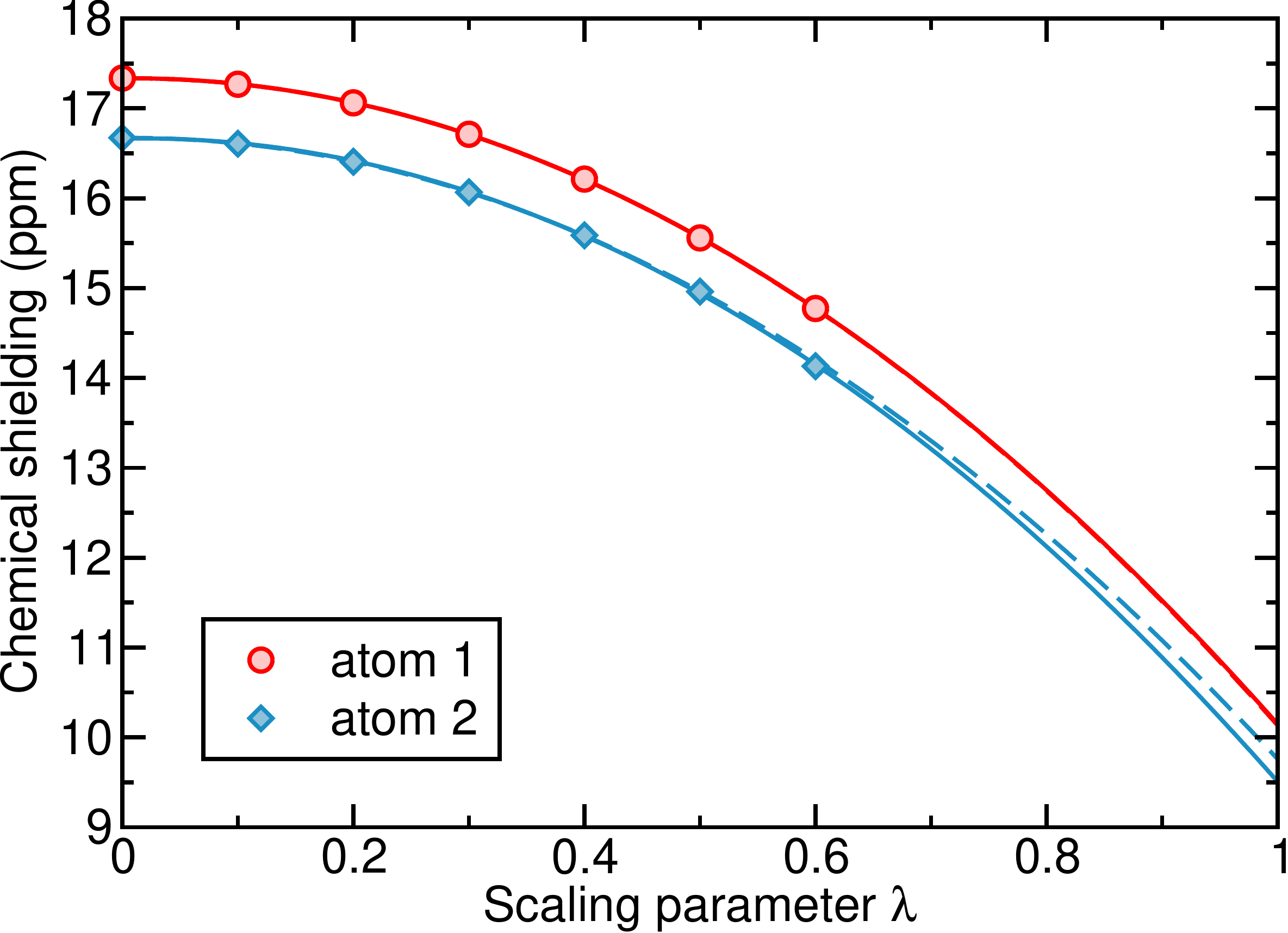}
  \caption{Isotropic chemical shielding $\frac{1}{3}\mathrm{Tr}(\bm{\sigma})$ as a function of the scaling parameter $\lambda$ for two distinct atoms in the $C2/c$ structure at $150$\,GPa. The red circles and blue diamonds represent the raw data, while the solid and dashed lines represent fits to a quadratic function used to extrapolate to $\lambda=1$. The solid line fits use all data points in the range $\lambda\in[0,0.6]$, while the dahsed line fits use only data points in the range $\lambda\in[0,0.5]$.}
      \label{fig:extrap}
\end{figure}
There is a subtle complication in the calculation of chemical shielding tensors renormalised by quantum motion in the case of hydrogen, which arises because some of the stochastic nuclear configurations distributed according to $|\chi(\mathbf{u})|^2$ exhibit metallic behaviour. This is a well-known artifact of first principles calculations of hydrogen\,\cite{ceperley_h_elph_coupling,azadi_gap_pimd} which is a consequence of giant electron-phonon coupling and the typical band gap underestimation of semilocal DFT, which should be removed as the experimental phases are insulating at the pressure range we consider\,\cite{gregoryanz_nature_phase_V,silvera_h2pre_science}, and the NMR spectroscopy signal in metals has additional contributions compared to insulators arising from the so-called Knight shift\,\cite{knight_shift}. To correct this artifact, we scale the Gaussian amplitude of the quantum zero-point wave function in Eq.\,(\ref{eq:gaussian}) as $s^2_{\mathbf{q}\nu}\to\lambda s^2_{\mathbf{q}\nu}$, and evaluate Eq.\,(\ref{eq:mc}) at multiple values of $\lambda$. We find that for $\lambda\leq0.6$ all stochastic configurations are insulating, but starting at $\lambda=0.7$ some of the stochastic configurations are metallic. We therefore only use results for $\lambda\in[0,0.6]$, and fit a quadratic function $\bm{\sigma}_{\mathrm{ZP}}(\lambda)=\mathbf{a}_0+\mathbf{a}_1\lambda+\mathbf{a}_2\lambda^2$ in this range of $\lambda$, and then extrapolate to the $\bm{\sigma}_{ZP}(\lambda=1)$ value to obtain an estimate of the full quantum zero-point renormalised chemical shielding tensor. We show an example of such extrapolation procedure for the isotropic chemical shielding $\frac{1}{3}\mathrm{Tr}(\bm{\sigma})$ in Fig.\,\ref{fig:extrap}. The raw calculations are depicted by the circles and diamonds for two different atoms of the $C2/c$ structure at $150$\,GPa. The solid lines are the fitted quadratic function using all data points in the range $\lambda\in[0,0.6]$, while the dashed lines are the fitted quadratic function using only data points in the range $\lambda\in[0,0.5]$. The two fits for atom $1$ are indistinguishable and lead to an uncertainty in the extrapolated isotropic chemical shielding of less than $0.02$\,ppm. Atom $2$ exhibits the largest uncertainty in any extrapolation, of $0.26$\,ppm, but most atoms exhibit significantly smaller uncertanties in the extrapolation. We have further reduced the extrapolation uncertainty by treating each atom as independent in the fit, but subsequently averaging the extrapolated chemical shielding tensors over the atoms that are crystallographically equivalent and should therefore exhibit the same NMR response.

We have performed analogous calculations for the deuterium isotope, but the larger mass of deuterium and the associated smaller vibrational amplitudes mean that non-metallic configurations were available for $\lambda\in[0,0.7]$.

\begin{figure}
  \includegraphics[scale=0.36]{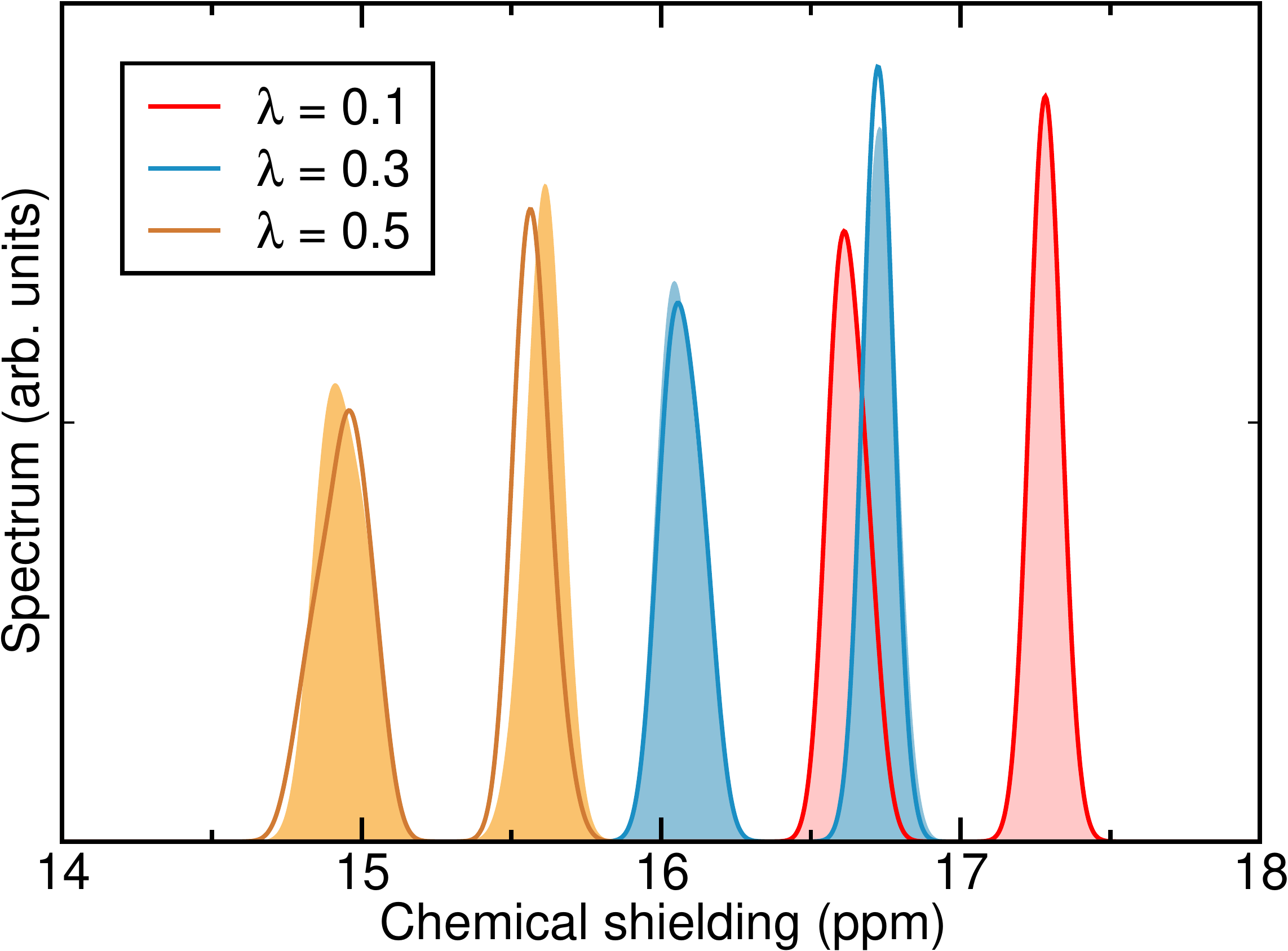}
  \caption{Isotropic chemical shielding $\frac{1}{3}\mathrm{Tr}(\bm{\sigma})$ at three different values of the scaling parameter $\lambda$ for the $C2/c$ structure at $150$\,GPa and using a Gaussian smearing of width $0.05$\,ppm. The solid lines depict the spectra obtained using a primitive cell and the filled distributions depict the spectra obtained using a $2\times1\times2$ supercell.}
      \label{fig:sc}
\end{figure}
The results presented so far have been obtained by considering quantum zero-point motion only in the atoms in the primitive cell. We expect that this is a good approximation for describing the vibrational effects on the chemical shielding tensor because this is a local quantity. To confirm this expectation, we have calculated the vibrational correction to the chemical shielding tensor in $C2/c$ at $150$\,GPa using a larger $2\times1\times2$ supercell containing $96$ atoms compared to the $24$ atoms in the primitive cell. The results are depicted in Fig.\,\ref{fig:sc} and confirm that the primitive cell calculations are well converged. We note that the results are shown for three values of $\lambda$ for which all configurations are insulating for both supercell sizes.

\section{Quadrupole interaction in $^2$H}

Deuterium exhibits a quadrupole moment which interacts with the electric field gradient at the nucleus, leading to an additional contribution to the NMR signal. This interaction depends on the so-called electric field gradient tensor $\bm{V}$, which characterises the quadrupolar interaction magnitude $C_q=eqV_{33}/h$ where $e$ is the elemental charge, $q$ is the nuclear charge, $V_{33}$ is the largest magnitude principal component of $\bm{V}$, and $h$ is Planck's constant. The $\bm{V}$ tensor is also characterised by the so-called asymmetry $\eta_q=(V_{11}-V_{22})/V_{33}$ with $0<\eta_q<1$. In Tables \ref{tab:efg-p21c},\ref{tab:efg-p63m}, \ref{tab:efg-c2c}, \ref{tab:efg-p6122}, \ref{tab:efg-pc}, and \ref{tab:efg-pca21} we report both $C_q$ and $\eta_q$ for the $P2_1c$, $P6_3m$, $C2/c$, and $P6_122$ structures at $150$\,GPa and for the $Pc$ and $Pca2_1$ structures at $250$\,GPa using the PBE functional.

\begin{table}
  \setlength{\tabcolsep}{6pt} 
  \caption{Electric field gradient $C_q$ and $\eta_q$ for the $P2_1c$ structure at $150$\,GPa using the PBE functional.}
  \label{tab:efg-p21c}
  \begin{ruledtabular}
  \begin{tabular}{ccc}
   Atomic site & $C_q$ (MHz) & $\eta_q$ \\
  \hline
  $1$ & $2.067\times10^{-1}$ & $0.05$ \\
  $2$ & $2.116\times10^{-1}$ & $0.05$ \\
  $3$ & $2.063\times10^{-1}$ & $0.02$ \\
  $4$ & $2.080\times10^{-1}$ & $0.02$ \\
  $5$ & $2.105\times10^{-1}$ & $0.04$ \\
  $6$ & $2.096\times10^{-1}$ & $0.01$ \\
\end{tabular}
\end{ruledtabular}
\end{table}

\begin{table}
  \setlength{\tabcolsep}{6pt} 
  \caption{Electric field gradient $C_q$ and $\eta_q$ for the $P6_3m$ structure at $150$\,GPa using the PBE functional.}
  \label{tab:efg-p63m}
  \begin{ruledtabular}
  \begin{tabular}{ccc}
   Atomic site & $C_q$ (MHz) & $\eta_q$ \\
  \hline
  $1$ & $2.056\times10^{-1}$ & $0.02$ \\
  $2$ & $2.145\times10^{-1}$ & $0.03$ \\
  $3$ & $2.178\times10^{-1}$ & $0.00$ \\
\end{tabular}
\end{ruledtabular}
\end{table}

\begin{table}
  \setlength{\tabcolsep}{6pt} 
  \caption{Electric field gradient $C_q$ and $\eta_q$ for the $C2/c$ structure at $150$\,GPa using the PBE functional.}
  \label{tab:efg-c2c}
  \begin{ruledtabular}
  \begin{tabular}{ccc}
   Atomic site & $C_q$ (MHz) & $\eta_q$ \\
  \hline
  $1$ & $1.944\times10^{-1}$ & $0.01$ \\
  $2$ & $2.005\times10^{-1}$ & $0.09$ \\
  $3$ & $2.033\times10^{-1}$ & $0.13$ \\
  $4$ & $1.995\times10^{-1}$ & $0.01$ \\
  $5$ & $1.933\times10^{-1}$ & $0.01$ \\
  $6$ & $2.005\times10^{-1}$ & $0.09$ \\
\end{tabular}
\end{ruledtabular}
\end{table}

\begin{table}
  \setlength{\tabcolsep}{6pt} 
  \caption{Electric field gradient $C_q$ and $\eta_q$ for the $P6_122$ structure at $150$\,GPa using the PBE functional.}
  \label{tab:efg-p6122}
  \begin{ruledtabular}
  \begin{tabular}{ccc}
   Atomic site & $C_q$ (MHz) & $\eta_q$ \\
  \hline
  $1$ & $2.087\times10^{-1}$ & $0.19$ \\
  $2$ & $1.945\times10^{-1}$ & $0.01$ \\
  $3$ & $2.030\times10^{-1}$ & $0.10$ \\
  $4$ & $2.014\times10^{-1}$ & $0.12$ \\
  $5$ & $2.022\times10^{-1}$ & $0.01$ \\
  $6$ & $2.050\times10^{-1}$ & $0.14$ \\
\end{tabular}
\end{ruledtabular}
\end{table}

\begin{table}
  \setlength{\tabcolsep}{6pt} 
  \caption{Electric field gradient $C_q$ and $\eta_q$ for the $Pc$ structure at $250$\,GPa using the PBE functional.}
  \label{tab:efg-pc}
  \begin{ruledtabular}
  \begin{tabular}{ccc}
   Atomic site & $C_q$ (MHz) & $\eta_q$ \\
  \hline
  $1$  & $2.113\times10^{-1}$ & $0.06$ \\
  $2$  & $2.131\times10^{-1}$ & $0.09$ \\
  $3$  & $2.118\times10^{-1}$ & $0.09$ \\
  $4$  & $1.337\times10^{-1}$ & $0.23$ \\
  $5$  & $1.162\times10^{-1}$ & $0.50$ \\
  $6$  & $1.310\times10^{-1}$ & $0.25$ \\
  $7$  & $2.069\times10^{-1}$ & $0.09$ \\
  $8$  & $2.119\times10^{-1}$ & $0.08$ \\
  $9$  & $2.172\times10^{-1}$ & $0.08$ \\
  $10$ & $1.237\times10^{-1}$ & $0.41$ \\
  $11$ & $1.414\times10^{-1}$ & $0.25$ \\
  $12$ & $1.266\times10^{-1}$ & $0.41$ \\
  $13$ & $2.104\times10^{-1}$ & $0.05$ \\
  $14$ & $2.159\times10^{-1}$ & $0.09$ \\
  $15$ & $2.134\times10^{-1}$ & $0.09$ \\
  $16$ & $1.172\times10^{-1}$ & $0.52$ \\
  $17$ & $1.307\times10^{-1}$ & $0.25$ \\
  $18$ & $1.145\times10^{-1}$ & $0.54$ \\
  $19$ & $2.076\times10^{-1}$ & $0.09$ \\
  $20$ & $2.103\times10^{-1}$ & $0.07$ \\
  $21$ & $2.162\times10^{-1}$ & $0.06$ \\
  $22$ & $1.376\times10^{-1}$ & $0.26$ \\
  $23$ & $1.229\times10^{-1}$ & $0.40$ \\
  $24$ & $1.388\times10^{-1}$ & $0.26$ \\
\end{tabular}
\end{ruledtabular}
\end{table}

\begin{table}
  \setlength{\tabcolsep}{6pt} 
  \caption{Electric field gradient $C_q$ and $\eta_q$ for the $Pca2_1$ structure at $250$\,GPa using the PBE functional.}
  \label{tab:efg-pca21}
  \begin{ruledtabular}
  \begin{tabular}{ccc}
   Atomic site & $C_q$ (MHz) & $\eta_q$ \\
  \hline
  $1$  & $2.100\times10^{-1}$ & $0.10$ \\
  $2$  & $2.135\times10^{-1}$ & $0.11$ \\
  $3$  & $2.207\times10^{-1}$ & $0.11$ \\
  $4$  & $1.079\times10^{-1}$ & $0.55$ \\
  $5$  & $1.106\times10^{-1}$ & $0.53$ \\
  $6$  & $1.089\times10^{-1}$ & $0.59$ \\
  $7$  & $2.101\times10^{-1}$ & $0.07$ \\
  $8$  & $2.183\times10^{-1}$ & $0.11$ \\
  $9$  & $2.161\times10^{-1}$ & $0.08$ \\
  $10$ & $1.080\times10^{-1}$ & $0.58$ \\
  $11$ & $1.079\times10^{-1}$ & $0.56$ \\
  $12$ & $1.091\times10^{-1}$ & $0.56$ \\
\end{tabular}
\end{ruledtabular}
\end{table}

\end{document}